\documentclass[prX,preprint]{revtex4-1}
\usepackage[english]{babel}
\usepackage[T1]{fontenc}
\usepackage{amsmath}
\usepackage{graphicx}
\usepackage{amsfonts}
\usepackage{amssymb}
\usepackage{units}
\newcommand{\DoF}{DoF }

\begin{document}

\title{Micromagnetic study of flux-closure states in Fe dots using quantitative Lorentz Microscopy}



\author{Aur\'elien Masseboeuf}
\altaffiliation[New ad. : ]{CEMES, CNRS - 31400 Toulouse - France}
\affiliation{CEA-INAC/UJF-Grenoble1 UMR-E, SP2M, Minatec Grenoble, F-38054}
\author{Olivier Fruchart}
\author{Fabien Cheynis}
\altaffiliation[New ad. : ]{CINaM, CNRS - 13288 Marseille Cedex 09 - France}
\author{Nicolas Rougemaille}
\affiliation{Institut N\'EEL, CNRS \& Universit\'e Joseph Fourier - BP166 - F-38042 Grenoble Cedex 9 - France}
\author{Jean-Christophe Toussaint}
\affiliation{Institut N\'EEL, CNRS \& Universit\'e Joseph Fourier - BP166 - F-38042 Grenoble Cedex 9 - France\\Grenoble INP - Grenoble, France}
\author{Alain Marty}
\author{Pascale Bayle-Guillemaud}
\affiliation{CEA-INAC/UJF-Grenoble 1 UMR-E, SP2M, Minatec Grenoble, F-38054}

\begin{abstract}
A micromagnetic study of epitaxial micron-sized iron dots is reported through the analysis of Fresnel contrast in Lorentz Microscopy. Their use is reviewed and developed through analysis of various magnetic structures in such dots. Simple Landau configuration is used to investigate various aspects of asymmetric Bloch domain walls. The experimental width of such a complex wall is first derived and its value is discussed with the help of micromagnetic simulations. Combination of these two approaches enables us to define what is really extracted when estimating asymmetric wall width in Lorentz Microscopy.  Moreover, quantitative data on the magnetization inside the dot is retrieved using phase retrieval as well as new informations on the degrees of freedom of such walls. Finally, it is shown how the existence and the propagation of a surface vortex can be characterized and monitored. This demonstrates the ability to reach a magnetic sensitivity a priori hidden in Fresnel contrast, based on an original image treatment and backed-up by the evaluation of contrasts obtained from micromagnetic simulations.

\end{abstract}

\maketitle

\section{Introduction}
The control of the motion of magnetic objects such as magnetic domain walls (DWs) and magnetic vortices is of great interest for their potential use in solid-state magnetic random access memories (MRAM) \cite{Allwood2005,Parkin2008}. An intense activity is currently devoted to the fundamental understanding of DW motion driven by either magnetic field or spin-polarized current, with the technological aim and fundamental need for understanding how to reach high mobilities (high speed with low field or low current).\\
Understanding and controlling the motion of domain-walls and vortices first requires a good knowledge of their internal micromagnetic structure. This structure is associated with one or more degrees of freedom (\DoF). For instance the core of a magnetic vortex can exhibit a magnetization in an up or down state, that may be switched by a magnetic field \cite{Yamada2007} or spin-polarized currents \cite{Van2006}. 
The internal structure of vortices and DWs is best studied in dots displaying a flux-closure state, because it is stabilized in its center owing to the self dipolar field \cite{Schneider2002}. The core orientation of the magnetic vortex combined with the chirality (clockwise or anticlockwise) of the flux closure, define two \DoF, that can be considered as bits in terms of data storage. Many studies have been devoted to disks with these two \DoF \cite{Schneider2001,Kimura2005a,Zhong2009}. Recently elongated dots with three \DoF (two in the central Bloch domain wall, one in the chirality) were demonstrated, first in self-assembled dots \cite{Cheynis2009} then extended to spin-valve dots \cite{Huang2009}. \\
Owing to its high lateral resolution and video capture rate, Transmission Electron Microscopy (TEM - and its associated magnetic imaging technique Lorentz Microscopy - LTEM) is a powerful tool to scrutinize the inner structure of magnetic objects such as vortex arrays \cite{Tonomura2002a}, vortices \cite{Beleggia2002a} or Bloch lines \cite{Masseboeuf2009APL}. The resolution capacities below \unit[10]{nm} \cite{Tanase2009} associated to a bulk magnetic sensitivity are of great interest for such fine analysis. Furthermore, ~\textit{in-situ} experiments (few tens of millisecond temporal resolution) can be carried out to monitor in real time these magnetic objects.\\
The purpose of the present manuscript is to analyse flux-closure states in micron-sized self-assembled dots via Lorentz microscopy, both under static conditions and while monitoring the quasi-static switching of internal degrees of freedom of the DW. It is illustrated how advanced image processing of experimental data combined with post-processing of micromagnetic calculations are crucial in getting the highest possible resolution and information out of experimental data. Moreover, such micromagnetic knowledge of the DW is used to analyse the sensitivity of Lorentz microscopy.
Section II describes the system under study and the experimental set-up. Section III is devoted to a simple analysis of Fresnel contrast used to retrieve quantitative information on the magnetic width of the DW and its comparison to micromagnetic modelling. Section IV deals with a detailed analysis of the phase retrieval approach based on Fresnel contrast to retrieve quantitative informations on both integrated magnetic induction and domain wall width. New possibilities offered by high-resolution Fresnel contrast analysis are illustrated in the last section by the real-time monitoring of a magnetization process inside the DW itself, based on the propagation of a surface vortex of diameter roughly \unit[10]{nm}.

\begin{figure}[h]
\includegraphics[width = \columnwidth]{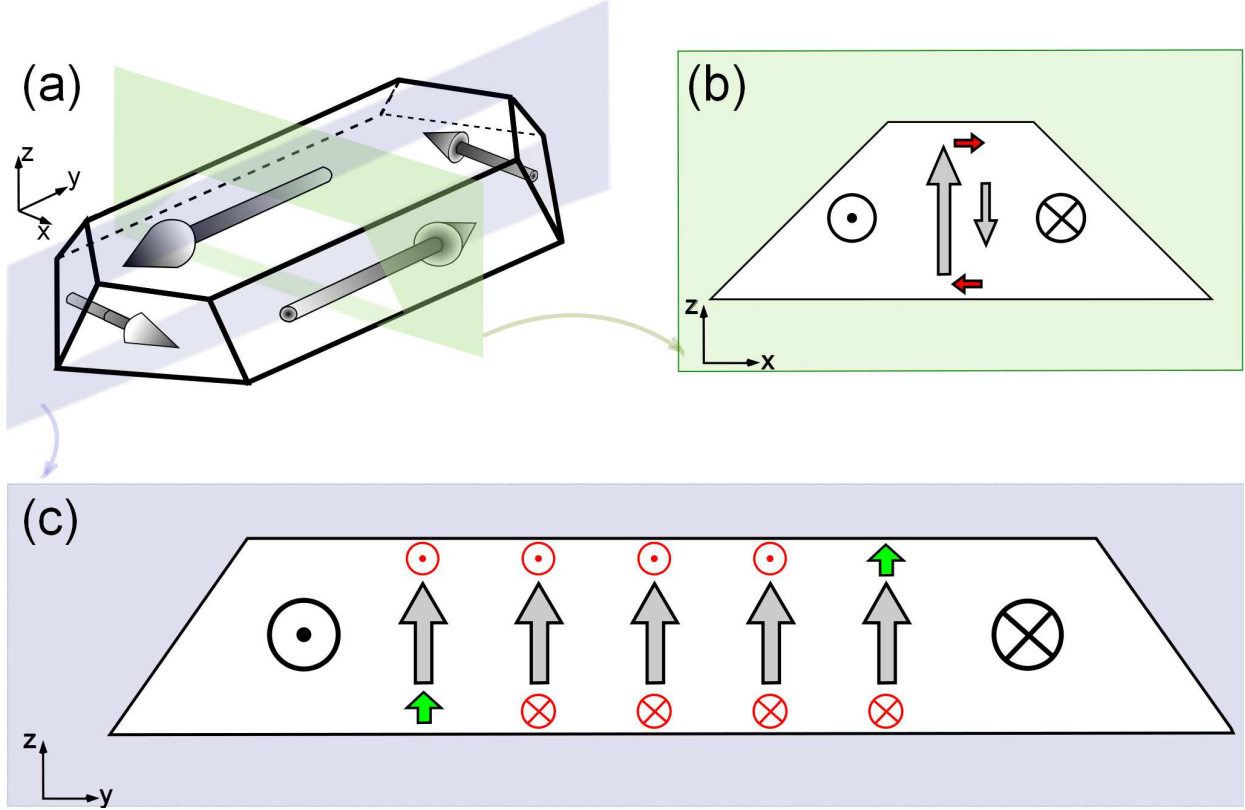}
\caption{\label{DotSimu} (Colour online) Scheme of the magnetization distribution in a single-crystalline elongated and faceted iron (110) dot. (a) Overview of the flux closure distribution in the dot. The two coloured planes are referring to the two following views. (b) Transverse section of the dot. The magnetic configuration of the asymmetric Bloch wall can be seen. The two N\'eel Caps are highlighted in red. (c) Longitudinal section of the dot. The overall finite Bloch wall is described here. Surface vortices are highlighted in green, as the two NCs are drawn in red.}
\end{figure}

\section{Experimental setup}
The nanostructures studied in this paper are micron-sized iron Fe(110) dots, synthesized using Pulsed-Laser Deposition under Ultra-High Vacuum conditions. The supporting surface is a \unit[10]{nm}-thick W layer epitaxially-grown on 350 micron-thick Sapphire (11$\bar{2}$0) wafers. These dots are faceted because they are single-crystalline and display low Miller indices crystallographic planes. Their elongated shape is due to the uniaxial anisotropy of the (110) surface. Details about the sample growth can be found elsewhere \cite{Fruchart2007}. Such dots have been extensively characterized during the past ten years by means of micromagnetic simulations \cite{Jubert2003,Fruchart2004}, MFM observations \cite{Jubert2003}, X-ray Magnetic Circular Dichroism Photo-Emission Electron Microscopy \cite{Hertel2005} and magnetotransport \cite{Cheynis2009b}. The micromagnetic structure of such dots is sketched in Figure \ref{DotSimu} and can be described as follows.\\
Above a lateral size of roughly \unit[250]{nm} and thickness larger than \unit[50]{nm} the magnetostatic energy of the dot is so large that it displays spontaneously a flux closure magnetic distribution (Fig. \ref{DotSimu}-(a)). Several types of flux-closure states exist in such dots \cite{Hertel1999,Bode2004}. The simplest of these is the combination of two main domains, antiparallel one to another and oriented along the long dimension of the dot. A Bloch wall of finite length and width lies at the boundary of these two domains (Fig. \ref{DotSimu}-(b)). Smaller domains oriented essentially along the in-plane short axis of the dot are located at its ends to enable the magnetic flux to close. 
The detailed inner structure of the Bloch wall is asymmetric \cite{Hubert1969,LaBonte1969}. It is composed of a main out-of-plane magnetization area and two opposites so-called N\'eel Caps (NCs) occurring at each surface of the dot with opposite in-plane magnetization (see Fig. \ref{DotSimu}-(b)) \cite{Foss1996}. 
At each end of the finite Bloch wall one finds a surface vortex enabling the magnetic flux to escape (in green on Fig. \ref{DotSimu}-(c)). These two vortices are unavoidable based on topological arguments for a flux-closure dot \cite{Arrott1979}.
Thus three \DoF exist in an elongated dot : the vertical polarity of the DW, the chirality of the flux-closure and the transverse polarity of Ncs couple (an information equivalent with the position of the two surface vortices). The controlled magnetic switching of this third \DoF was demonstrated recently \cite{Cheynis2009, Cheynis2009b}.

Micromagnetic simulations were performed using a custom-developed code based on a finite-differences scheme (prismatic cells) \cite{Fruchart2004}.
Here both a 3D and a 2D version of the code were used. The former permits the accurate description of the complex magnetic structure arising in three-dimensional structures, while the latter allows to address simple cases such as an infinitely-long domain wall, and thus extract the essentials of the physics at play. We used the bulk magnetic parameters for Fe : exchange $A=\unit[2\times10^{-11}]{J/m}$, fourth-order magnetic anisotropy $K=\unit[4.8\times10^4]{J/m^3}$, and magnetization $M_{\mathrm{s}}=\unit[1.73\times10^6]{A/m}$.

Two microscopes were used for Lorentz Microscopy : a JEOL 3010 with a thermionic electron gun and a FEI Titan fitted in with a Schottky gun. Both of them are working at 300kV and are fitted with a Gatan Imaging Filter for zero loss filtering \cite{Dooley1997} and thickness mapping \cite{Malis1988}. The Titan is also equipped with a dedicated Lorentz lens for high resolution magnetic field-free imaging while the JEOL is fitted with a conventional objective mini-lens initially dedicated to low magnification imaging. In-situ experiments were performed using the field produced by the objective lens of the microscope (previously calibrated using dedicated sample holders mounted with a Hall probe). The sample was then tilted to produce an in-plane magnetic field. Magnetic field values provided in that manuscript refer to the in-plane component of the field with respect to the tilt angle. Sample was prepared using a mechanical polishing and ion milling. Phase retrieval using the Transport of Intensity equation \cite{Paganin1998} was thus coupled to substrate contribution removal as proposed in \cite{Masseboeuf2009Ultra}.

\begin{figure}[h]
\includegraphics[width = \columnwidth]{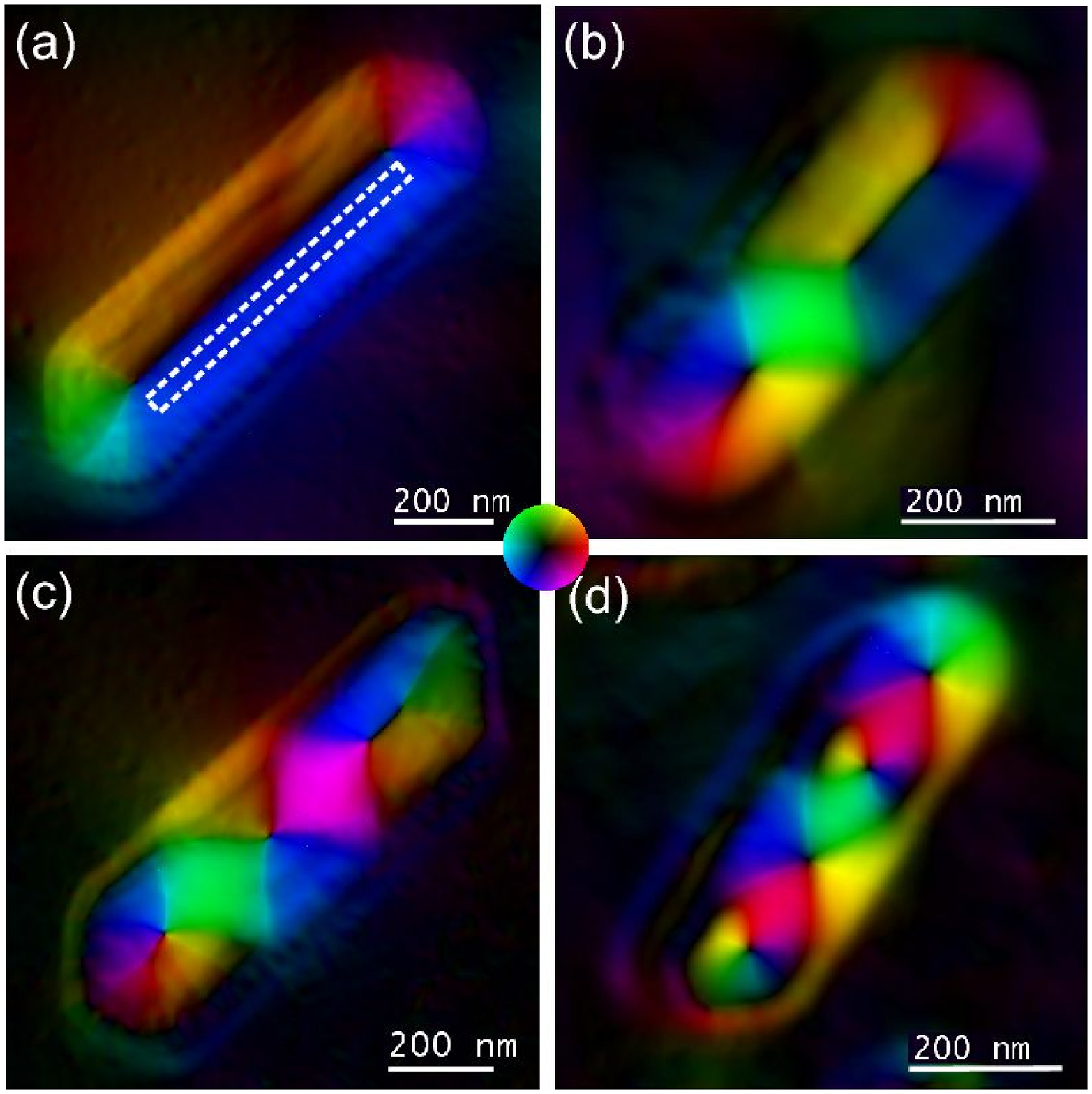}
\caption{\label{FigTIEFedot} (Colour online) Four different magnetic configurations of the iron dots. \textbf{(a)} Landau state, as described in the second section. The dashed area corresponds to the location where quantitative data where extracted. \textbf{(b)} 2-Landau state where two Bloch walls and thus two opposite chiralities are found in the same dot. \textbf{(c)} and \textbf{(d)} 3 and 4-Landau states where most of the domain walls collapsed into a vortex state due to their limited length. The colour disc at the center of the images displays the direction (hue) and the intensity (brightness) of the integrated magnetic induction used in the maps. The defocus value used for the reconstruction is around \unit[250]{$\mu m$}.}
\end{figure}

\begin{figure}[h]
\includegraphics[width = \columnwidth]{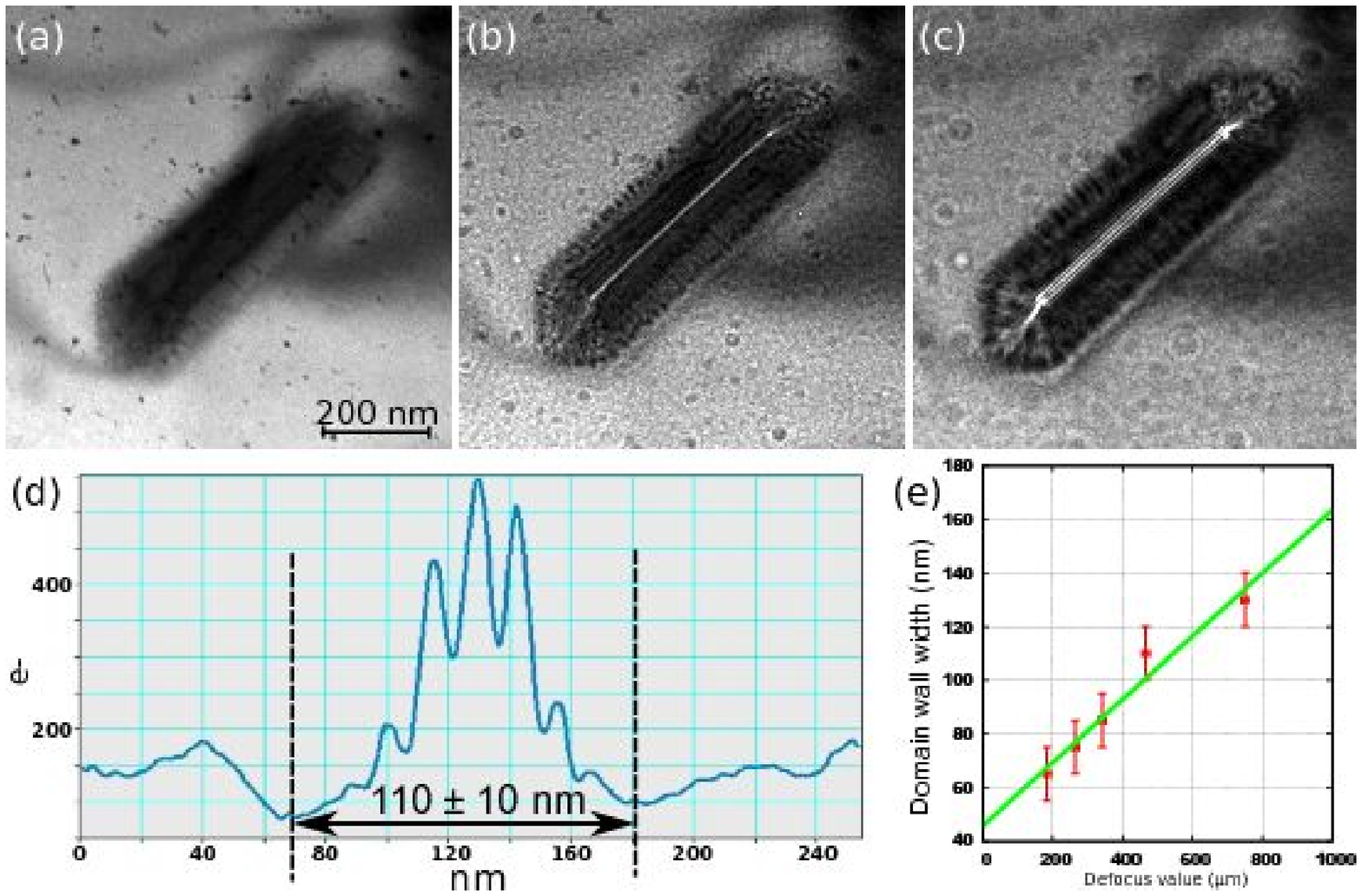}
\caption{\label{DWwidthExp} (Colour online) (a-c) Images of a Fe (110) dot for different defocus values (values of defocus are 0, 200 and $\unit[500]{\mu m}$ for (a) to (c) respectively). The thickness of the dot was estimated (see in the text) to \unit[90]{nm}. (d) Intensity profile of the convergent wall obtained from (c) and associated width measurement. (e) Zero-defocus approximation for domain wall width measurement.}
\end{figure}

All observations presented here are based on Fresnel contrast \cite{Chapman1984} of LTEM. Considering geometrical optics, its formation results from the overlap of two parts of the electron beam experiencing two different Lorentz forces. An image formed slightly over- or underfocused then results in bright or dark lines, highlighting the domain walls position. In the case of a coherent electron source where electrons have to be described as waves and no more as particles, the contrast give rise to interference patterns in overlapping areas (such images are subsequently often denoted \textit{in-line} holograms). This interference pattern contains further informations about the DW inner structure as it will be explained in the last section.\\
The sample geometry was chosen to reveal the Bloch wall contrast in Fresnel images (see Fig.\ref{DWwidthExp}). Electron are thus crossing the sample perpendicularly to the magnetic domains of the dot and parallelly to the Bloch wall magnetization. Fig. \ref{FigTIEFedot} presents four different flux-closure states. These maps were built using the phase retrieval approach of the Transport of Intensity Equation \cite{Paganin1998} linking the phase gradient in the observation plane to the intensity variation along the optical axis. The reconstructed phase contains an electrostatic and a magnetic component \cite{Aharonov1959} that we discuss hereafter. Due to the observed flux closure states we reliably make the approximation that the magnetic signal can be associated to the integrated magnetization inside the dots thus neglecting any significant demagnetizing field.\\
To retrieve quantitative magnetic information by suppressing the electrostatic contribution several techniques are known and are reviewed in \cite{Dunin-Borkowski2004}. The first technique we used is described in \cite{Masseboeuf2009Ultra} and enables to get rid of the electrostatic contribution of the substrate considering a constant gradient of substrate thickness. A value of \unit[150$\pm$50]{nm.T} for the integrated magnetic induction was found where the two surfaces of the dot are parallel and the electrostatic contribution of the dot vanishes (i.e. between the facets and the Bloch wall - see dashed area in Fig. \ref{FigTIEFedot}-(a)). Considering an experimental thickness of \unit[70]{nm} (estimated with the log-ratio technique \cite{Malis1988} using a value of the inelastic mean free path of \unit[80]{nm} for iron at \unit[300]{kV}), this value is in good agreement with the bulk iron saturated magnetization ($\mu_0 M_{\mathrm{s}}=\unit[2.17]{T}$). On the other hand, a second method is to reverse the chirality of the dot by an applied field \cite{PapierChiralite}. We perform a subtraction of two phase shifts with an opposite magnetic contribution but the same electrostatic contribution. In that case both contributions are differentiated. A value of \unit[140$\pm$50]{nm.T} was found for the integrated magnetization in the dot and a thickness of \unit[80$\pm$20]{nm} was also confirmed analysing the electrostatic part (assuming a \unit[22]{V} potential for iron). Eventually, a third procedure using a 180$^{\circ}$ reversal of the TEM sample (outside of the microscope) was used leading to the same value. Such method is more convenient to use as it is possible to analyse more complex magnetic structures. Analysing the central part of the diamond configuration (Fig. \ref{FigTIEFedot}-(b)) where the dot is undoubtedly uniformly magnetized could thus be carried out to confirm our previous results. An integrated magnetization of \unit[220$\pm$50]{nm.T} was found which confirms our previous measurements. Giving an experimental thickness of \unit[100]{nm} finally leads to an experimental value of the magnetization in iron : $M_{\mathrm{s}}=\unit[1.7 \pm 0.4 \times10^6]{A/m}$.\\

Other magnetic distributions (Fig. \ref{FigTIEFedot}-(b-d)) can be viewed as double, triple and quadruple Landau (or also under the generic name of diamond states \cite{Jubert2003}), the prefix referring to the number of Bloch walls or vortices (if the domain wall collapses due to a too small length) inside the dot. Any of these configurations may be prepared regarding the shape of the dot and its magnetic history. It may be used to favour the occurrence of one or another type of state. As a general rule a saturation magnetic field (between 2 and \unit[3]{T}) applied perfectly perpendicular to the dot (along z) yields a Landau state, whereas higher order states are obtained upon applying the field with a combination of an azimuthal and polar angle. Such multi-walled structure can be of fundamental use for wall length study as demonstrated in \cite{Masseboeuf2010PRL}.

\section{Fresnel contrast analysis}
Fig. \ref{DWwidthExp} displays three images taken for different defocus values. The 180$^{\circ}$ Bloch wall induces a bright line in the middle of the dot. At both extremities of this line weaker bright lines emerge due to N\'eel walls (more easily seen on high defocused images as Fig. \ref{DWwidthExp}-c). The brighter spots found at each extremity of the Bloch DW arises from the locally high curl of magnetization and are not linked to surface vortices presented in Fig. \ref{DotSimu}-c.\\
This contrast can be used to assess the width of the Bloch wall. We used the zero-defocus approximation that consists of a linear regression for a focal series of domain wall contrast width measurement \cite{Lloyd2002}. This \textit{old fashioned} \cite{Warrington1964} method has been widely used, commented and criticized in the past. We rely here on the validity of the method regarding the large width (well above \unit[10]{nm}) \cite{Gong1987} and asymmetry \cite{Marti1977} of the domain walls studied in the present work. Moreover, the main goal here is to compare the measured width with a real micromagnetic case to understand the meaning of such a method. The width of the convergent wall (\textit{i.e.} the width measured in Fresnel contrast) is estimated by taking the width of the interference pattern, namely the distance between the two outermost bright fringes (see Fig. \ref{DWwidthExp}-(d)). The extreme bright fringes are chosen when there are visible on both side of the main center bright line and when their intensity is more than 5 \% the center fringe intensity. We found a value of \unit[45$\pm$5]{nm} for the asymmetric Bloch wall width.\\

\begin{figure}[h]
\includegraphics[width = \columnwidth]{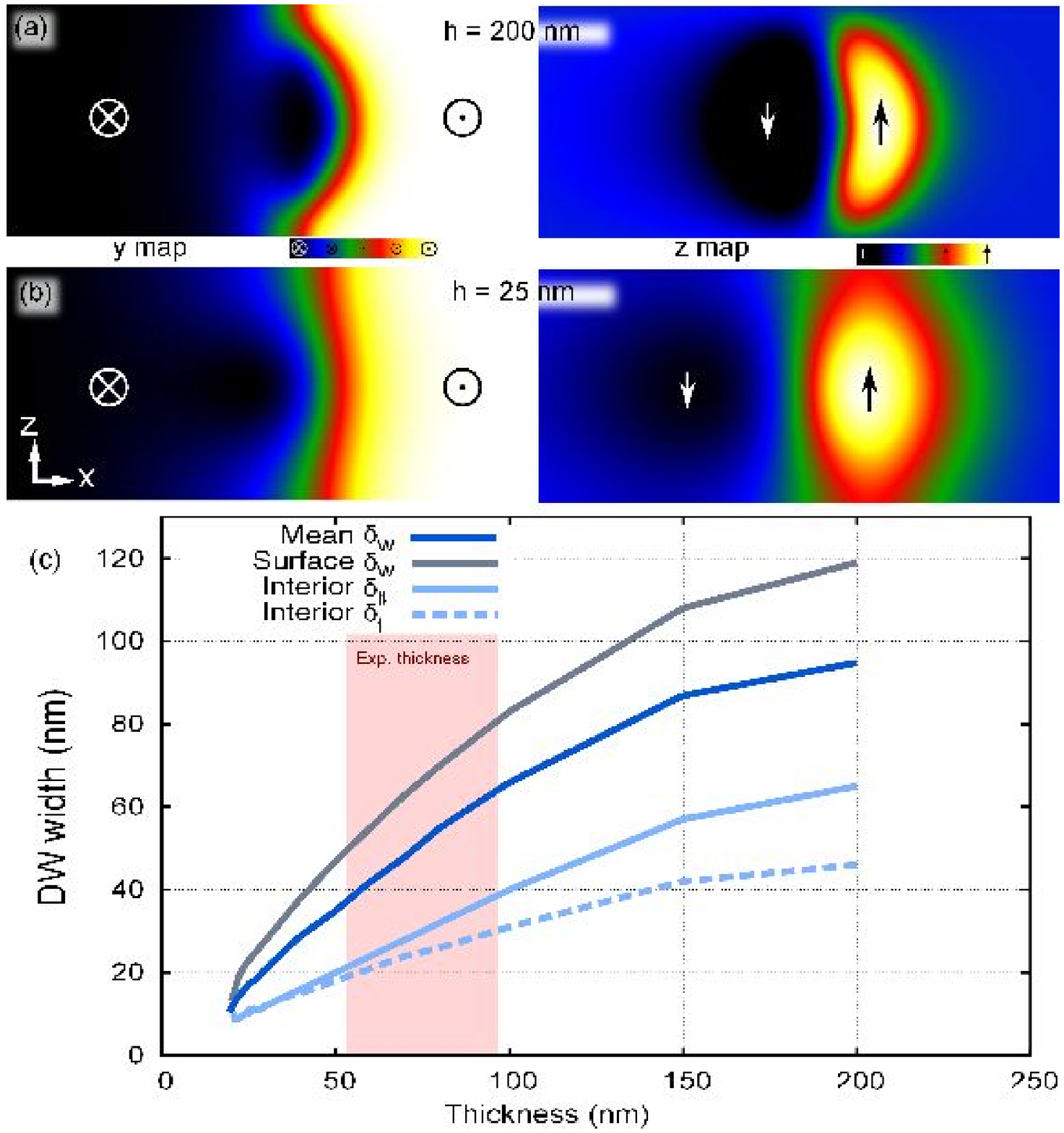}
\caption{\label{DWwidth} (Colour online) Domain wall width measure with respect to the thickness on simulated iron bar. \textbf{(a)} y- and z-components of the magnetization on a wall profile for a \unit[200]{nm}-thick iron bar. \textbf{(b)} Same as in (a) for a \unit[25]{nm}-thick iron bar with same aspect ratio. \textbf{(c)} Plot of the wall width versus thickness using various description of the domain wall width (see text). Experimental thicknesses area is reported on the graph.}
\end{figure}

To understand the signification of this value, we used our 2D micromagnetic code considering infinitely long (in the $y$ direction) iron (110) bars with a thickness over width ratio of 0.2 which is a typical experimental value. Due firstly to the four-fold magnetocrystalline anisotropy, and secondly to the three-dimensional nature of the system, the domain wall width cannot be easily defined by a tangent like in the text-book case \cite{Bloch1932}. Choosing the most suitable definition of a DW width is mandatory for the analysis of both experiments and simulations. We decided to use the formula described e.g. by Hubert \cite{Hubert1998} integrating over the $\theta$ angle corresponding to the magnetization orientation with respect to the wall direction (y) (both magnetization variations along $x$ and $z$ are described in Fig. \ref{DWwidth}-(a-b)):
\begin{equation}
\label{EqDW}
 \delta_w = \int_0^{\infty} \sin \theta_x dx
\end{equation}
This formula enables a clear bounding of the domain wall with two zeros values. Nevertheless it still can be integrated or averaged over the thickness of the dot (\textit{i.e.} along the $z$ direction), or applied at any height, the surface and middle-height being of particular interest. Fig. \ref{DWwidth}-(c) presents resulting values of width measurements considering various descriptions given hereafter. Several comments can be made on these results. In the low thickness regime, for any definition the DW width fits roughly linearly with the thickness. In that case one often uses the name of \textsl{vortex wall} for the obvious reason of curling of magnetization in the (x,z) plane (Fig. \ref{DWwidth}-(b)). 
The DW width seems hardly to saturate for large thickness, whereas for \unit[200]{nm} a steady value was nearly reached by Rave and Hubert \cite{Rave1998}. This probably because they used a much larger uniaxial anisotropy value than the one in Fe, yielding narrower domain walls. This steady width is much smaller than \unit[200]{nm} in their case, so that bulk properties are already reached. In our case the bulk wall width is larger and the DW is still geometrically constrained at \unit[200]{nm} thickness and ever more below (Fig. \ref{DWwidth}-(a)).

Aside the DW, we observe an area with a small volume of vertical magnetization with a sign opposite to that of the core of the DW (see z-maps on Fig. \ref{DWwidth}-(a-b)). Whereas this was already visible in the early simulations of Hubert \cite{Hubert1969} and LaBonte \cite{LaBonte1969}, this antiparallel volume is nearly absent in the extensive calculations reported more recently because again of the choice of a high value of anisotropy \cite{Korzunin2006,Korzunin2010}. The presence of this small volume implies a more careful description of how the domain wall width should be defined : with or without taking this volume into account. Two distinct approaches can be used considering ($\delta_{\uparrow\downarrow}$) or not ($\delta_{\uparrow}$) this opposite component of the wall (i.e. when $m_z$ $\leqslant$ 0, see also Fig. \ref{DotSimu}-(b)). As an illustration, Fig. \ref{DWwidth}-(c) shows the mean DW width computed from Eq. \ref{EqDW} and integrated over the thickness for both cases. The results show large differences with the full integral calculation showing that care should be taken when discussing the width of such walls. Experimentally it is found that for a thickness of \unit[70]{nm} the width is $\delta=\unit[45]{nm}$. This value fits better with the mean $\delta_{w}$ definition which is coherent with the integration along the electron path which is made when using Lorentz microscopy. Therefore such a description is not well suited to describe the width of complex asymmetric Bloch DWs. Nevertheless this measurement gives useful information on the wall profile as the inner width can be estimated with respect to the measured average value. We will see in the last part of that study how it is possible to analyse carefully Fresnel contrast to obtain more spatial information on the magnetization distribution in such walls.\\

Combination of experimental width measurement with micromagnetic simulation is thus mandatory to explain what is obtained during LTEM analysis. In the present work, an experimental quantitative value for the wall width was obtained but it was clear that neither a wall profile nor a wall description could be extracted by a simple contrast analysis. Micromagnetic simulation was thus an unavoidable tool we used to translate our findings in terms of magnetic length. Main result of such a work is that domain wall width measurement using LTEM cannot be reduced to an experimental snapshot and requires a good micromagnetic description (and not a simple text-book approximation) before conclusions can be made.\\

\begin{figure}[h]
\includegraphics[width = \columnwidth]{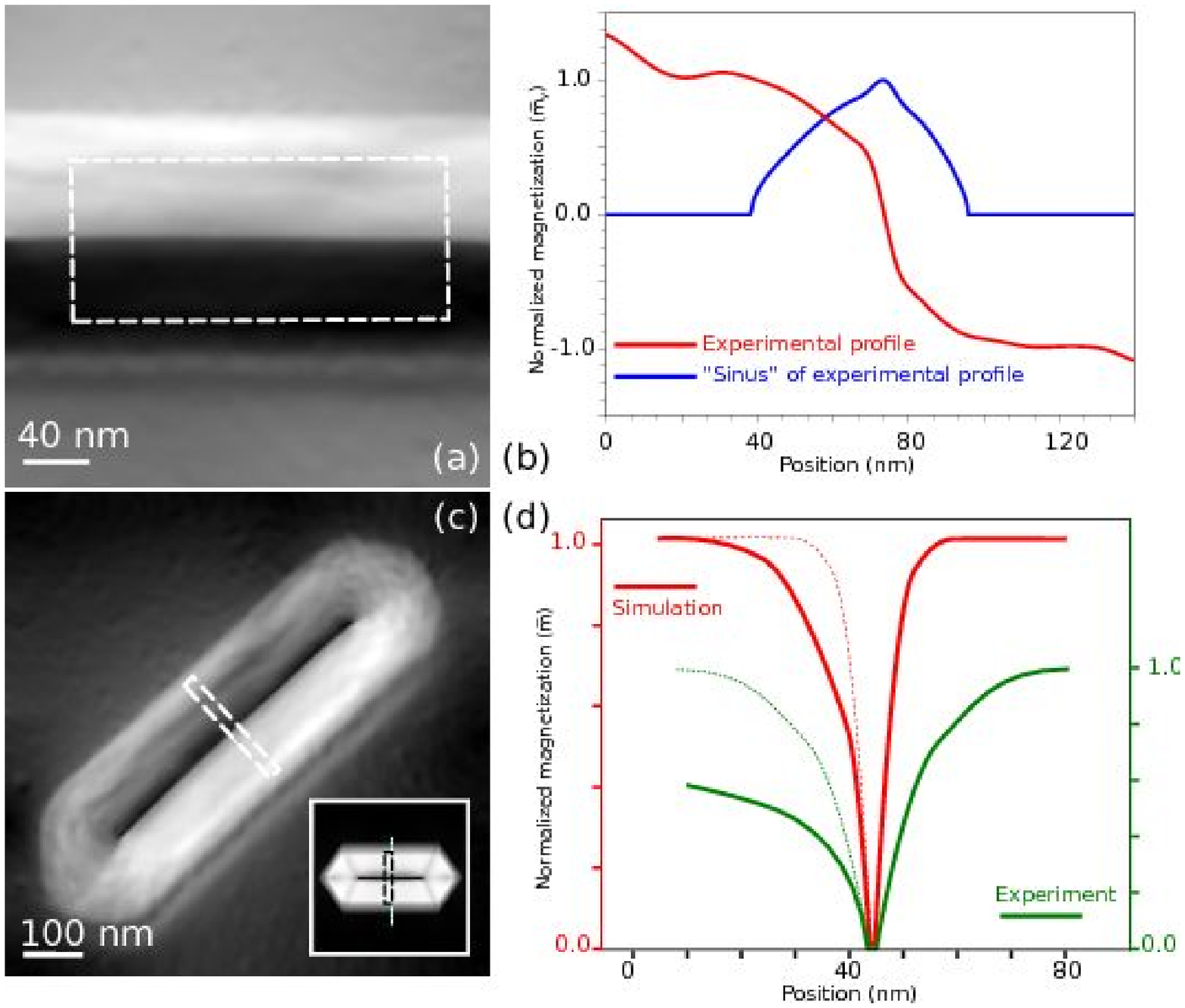}
\caption{\label{TIEanalysis}  (Colour online) TIE reconstruction analysis. (a) Experimental phase gradient along $x$ displaying the $y$ component of the integrated magnetization. (b) Experimental profile taken in (a) and its associated "positive sinus" approximation (see text). (c) Experimental modulus of the integrated induction calculated using two perpendicular phase gradients. The inset is a simulated integrated induction map from micromagnetic modelling. The dashed lines show localisation of profiles presented in (d). (d) Profiles extracted from (c). The continuous lines display the extracted profiles of the integrated magnetic induction maps for the simulated and experimental maps. Dashed lines show what would be expect for a symmetric wall.}
\end{figure}

\section{Phase-shift analysis}
As TIE approach is based on defocused images we did not expect a better resolution in measuring the domain wall width by such a method. However a phase shift is known to contain more information than a simple Fresnel image (which is definitely true regarding out-of-axis \textit{in-focus} holography, but impossible to implement in that case due to the large dimensions of the structures). We performed wall width measurements using a phase shift gradient along $x$ yielding to a $\bar{m_y} \cdot t$ map, $\bar{m_y}$ being the component along $y$ of the integrated magnetization and $t$ the local magnetic thickness. Such values were then normalized between +1 and -1 to approximate the cosine of the wall angle (see Fig. \ref{TIEanalysis}-(a)). We then estimate the sinus used in Eq. \ref{EqDW} by taking the real part of $\sqrt{1-(\bar{m_y}\cdot t)^2}$. An advantage of such description is to avoid experimental fluctuations around the maximum value of $\bar{m_y}\cdot t$. We finally found a value of $\unit[54\pm10]{nm}$ for domain wall width which is in accordance with the zero-defocus estimation. The small over-estimation could be explained by the use of an out-of-focus method and thus linked to the defocalisation value used in TIE method.\\
Fig. \ref{TIEanalysis}-(c-d) presents a map of $\bar{m}\cdot t = \sqrt{(\bar{m_x}\cdot t)^2+(\bar{m_y}\cdot t)^2}$. The map is obtained from two perpendicular phase gradients and contains a supplementary information on the third \DoF described above. This third \DoF is carried by the chirality of the asymmetric Bloch wall. Thus, $\bar{m} \cdot t$ decreases differently one side or another from the Bloch wall where $\bar{m} \cdot t$ vanishes due to perpendicular magnetization. The two NCs are perfectly antiparallel and give no signal in $\bar{m} \cdot t$ as they cancel one another. On the contrary, the presence of the small volume of vertical magnetization antiparallel to the wall magnetization (see Fig. \ref{DWwidth}-(a-b)) induces a local decrease of $\bar{m} \cdot t$. In both experimental (Fig. \ref{TIEanalysis}-(c)) and simulated (bottom right inset on same figure) cases, one can see an asymmetry in $\bar{m} \cdot t$ profile (Fig. \ref{TIEanalysis}-(d)). This asymmetry gives a unambiguous information on the position of the antiparallel magnetization volume of the asymmetric Bloch wall thus leading to the chirality of the wall. Adding the wall polarity information that was recently proposed in \cite{Ngo2011} for vortices could lead to a complete description of the three \DoF in such dots.

\begin{figure}[h]
\includegraphics[width = \columnwidth]{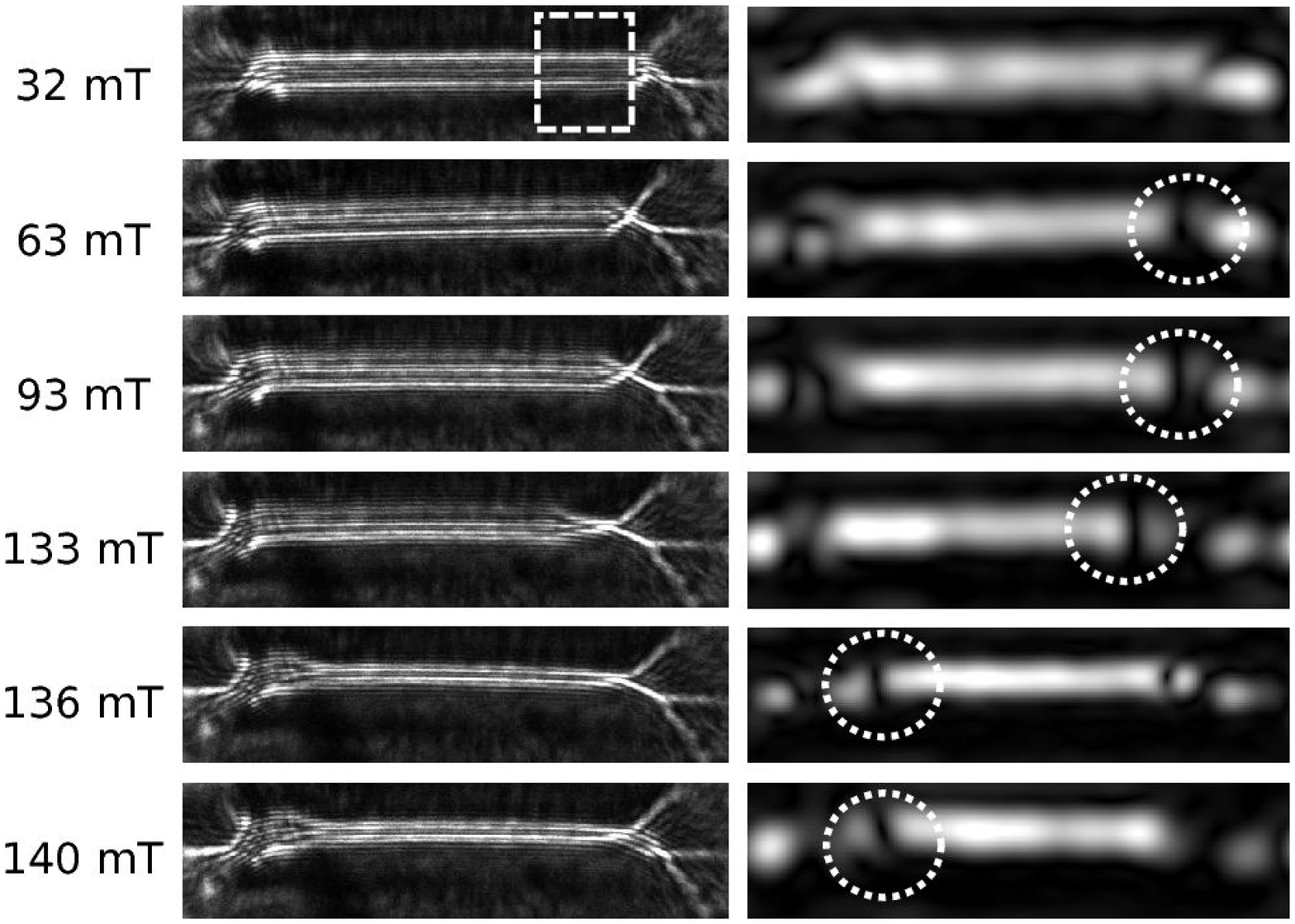}
\caption{\label{GPAdots}Experimental Fresnel contrast (left column) and associated amplitude (right column) analysis for corresponding  fringe frequency. The corresponding in-plane component of the applied magnetic field is indicated in the first column for each row. Dashed square on first image presents the zoomed part used in Fig. \ref{SwitchSimu}. Dashed circles on amplitude images are guide to the eyes to follow the surface vortex.}
\end{figure}

\section{Fresnel contrast exploration}
From a general point of view LTEM suffers from a limited sensitivity because an integration is made over the electron path. Here we demonstrate how the induced fringes of a convergent wall can be analysed to yield highly-resolved information about the domain walls. A comparison with micromagnetic modelling and contrast simulation is also given to confirm our observations.\\

We focus on the process of reversal of N\'eel caps, mediated by the motion of a surface vortex along the length of the domain wall. One can find detailed information on the process in \cite{Cheynis2009,Cheynis2009b}.
One considers in that section that the surface vortex is a simple perturbation of the Bloch wall. Its displacement along the Bloch wall only slightly modifies the local magnetization distribution. When increasing the defocus of the imaging lens, the overlap of the electrons coming from the neighbouring domains becomes wider and as a result the number of interference fringes increases. If the defocus value is high enough (namely close to a millimetre) the interference pattern can be compared to a small off-axis hologram \cite{Cowley1992}, bearing a wealth of information. Inner details of the DW can then be derived from the analysis of these fringes, such as the location of a surface vortex. The result is shown in Fig. \ref{GPAdots}.\\
Both original and amplitude images of the fringes are presented. The amplitude image corresponds to a wave reconstructed with the spatial frequency of the fringes and thus displays the location of the fringes in the image. The first image taken at \unit[32]{mT} is the starting point of the series with the vortex located to the right side. On the following images the fringe perturbation created by the vortex displacement can be seen with the straight dark line in the amplitude image (highlighted with dashed circles on Fig. \ref{GPAdots}) indicating that the fringes are locally suppressed.
The perturbation is moving from the right extremity (at 63, 93 and \unit[133]{mT}) along the domain wall and reaches its other extremity. The main transition occurs here between 133 and \unit[136]{mT}. After this transition (above 150 mT), the perturbation disappears and the wall exhibits a contrast similar to that observed on original images. The phenomenon is hysteretic and on the decreasing field series the magnetic switching happens between 123 and \unit[121]{mT}. We note here that an extra feature appears on the amplitude image at \unit[136]{mT} (dark mark on the right of the wall). We are confident in the fact that such mark is an artefact of our method because nothing can be seen in the fringes. Moreover such feature is absent on other images. This could be associated to the curl of magnetization around the Bloch wall which is slightly modified during magnetization process (the magnetization in the two main domains of the dot responds to the applied field) and which prevents from recording a prefect in-line hologram.\\

\begin{figure}[h]
\includegraphics[width = \columnwidth]{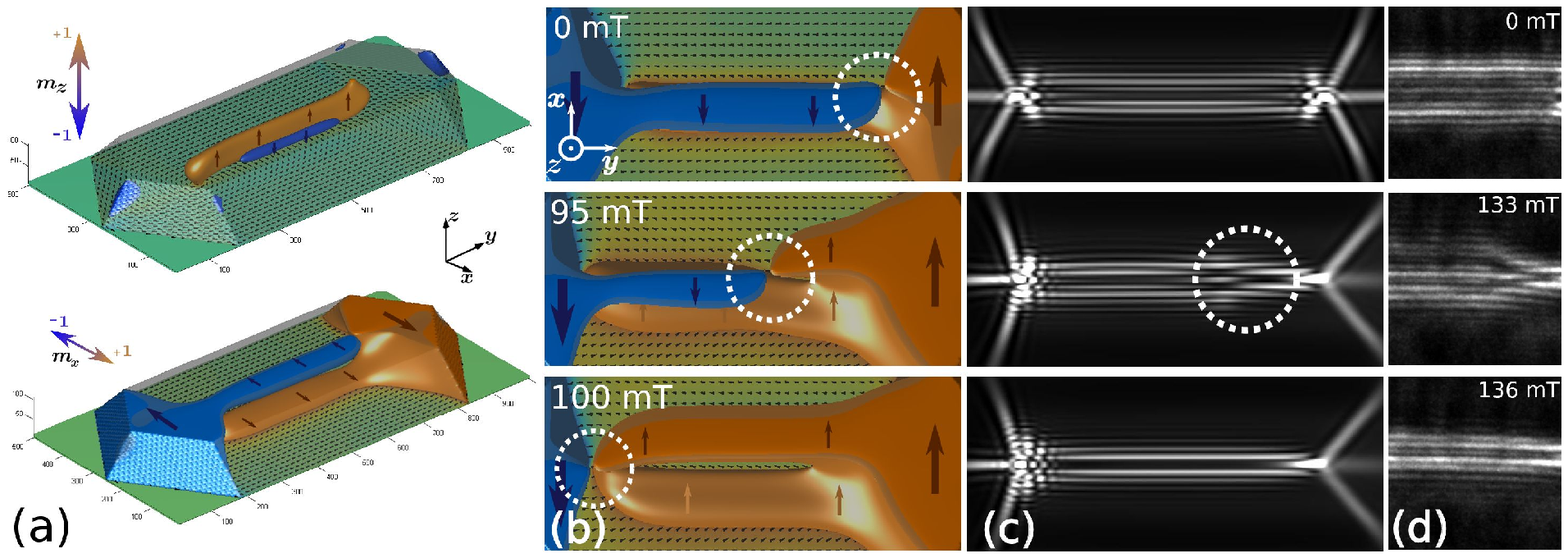}
\caption{\label{SwitchSimu} (Colour online) Three dimensional micromagnetic modelling of the iron dots during a transverse magnetic field application and associated computed Fresnel contrast. \textbf{(a)} The $z$ and $x$ components of the magnetization displayed with a minimum threshold of $\pm \frac{M_{\mathrm{s}}}{2}$  to provide a readable open-view of the magnetization inside the dot. $z$ map displays mainly the central wall and $y$ map exhibits clearly the NCs. \textbf{(b)} Three different modellings (x-component displayed as in (a)) for various transverse applied fields (along $x$). \textbf{(c)} Associated Fresnel contrast simulation for a defocus of \unit[750]{$\mu m$}. \textbf{(d)} Corresponding experimental Fresnel contrast extracted from Fig. \ref{GPAdots}. Values of experimental applied field are given.}
\end{figure}

A code to simulate Fresnel contrast from three-dimensional micromagnetic simulations was developed. The tilt of the sample used for in-plane field application is modelled using a barycentre approach : each three-dimensional voxel of the micromagnetic simulation is projected onto a plane. If the projected voxel is misaligned with the new mesh of the magnetic distribution, its value is spread on the four nearest neighbours depending on its center of mass in this square. 
The phase shift is associated with a simple object plane without a thickness. Note that the defocus used experimentally and in simulation is very important (close to mm) regarding the thickness of the dot (\unit[100]{nm}). The assumption of a simple phase object with no thickness should thus be valid.\\
Fig. \ref{SwitchSimu}-(a) and (b) present the three-dimensional modelling. Simulated Fresnel contrast obtained from simulation with various transverse applied fields (Fig. \ref{SwitchSimu}-(b-c)) yield a similar perturbation that on experimental Fresnel fringes (Fig. \ref{SwitchSimu}-(d)). We are therefore confident that such a perturbation in the fringe pattern is clearly related to the presence of a surface vortex.

\section{Conclusion}
We have described the micromagnetic configuration of epitaxially grown (110) iron dots. The micromagnetic knowledge of these dots has been retrieved by Lorentz microscopy observations. These observations enabled to measure the domain wall width of an asymmetric Bloch wall, and to compare it with micromagnetic simulations. Successful comparisons allowed to define more clearly how such asymmetric wall may be defined and what magnetic width measured by LTEM refers to. Two of the three degrees of freedom in such structures where also derived. Besides, various magnetic configurations were highlighted during phase retrieval processes. All these configurations could be easily obtained by modifying the magnetic history of the dots. Finally, the switching process of the N\'eel Caps surrounding the central Bloch wall was perfectly described with the used of interference patterns created in Fresnel contrast, underlining the potential of high-coherence microscopes in getting the highest resolution information using Lorentz microscopy. With such a detection in our LTEM measurement we could estimate the spatial resolution of LTEM to \unit[10]{nm} corresponding to the vortex diameter probed.

\bibliographystyle{unsrt}
\bibliography{UltraFedot}

\end{document}